\documentclass[onecolumn]{aastex631}
\pdfoutput=1 
\usepackage{amsmath,amstext}
\usepackage[T1]{fontenc}
\usepackage{apjfonts} 
\usepackage[figure,figure*]{hypcap}
\usepackage{CJKutf8}
\usepackage{longtable}
\usepackage[section]{placeins}

\usepackage{xspace}
\newcommand{\lido}{L\textit{i}DO\xspace}

\shortauthors{Pike et al.}

\begin{document}

\title{L$i$DO: Discovery of a 10:1 Resonator with a Novel Libration State}

\author[0000-0003-4797-5262]{Rosemary~E. Pike}
\affiliation{Center for Astrophysics | Harvard \& Smithsonian; 60 Garden Street, Cambridge, MA, 02138, USA}

\author[0000-0001-5061-0462]{Ruth Murray-Clay}
\affiliation{Astronomy and Astrophysics, University of California Santa Cruz}

\author[0000-0001-8736-236X]{Kathryn Volk}
\affiliation{Planetary Science Institute, 11700 East Fort Lowell, Suite 106, Tucson, AZ 85719, USA}

\author[0000-0003-4143-8589]{Mike~Alexandersen}
\affiliation{Center for Astrophysics | Harvard \& Smithsonian; 60 Garden Street, Cambridge, MA, 02138, USA}

\author[0009-0007-9017-4010]{Mark~Comte}
\affiliation{Campion College and the Department of Physics, University of Regina, Regina, SK S4S 0A2, Canada}

\author[0000-0001-5368-386X]{Samantha~M. Lawler}
\affiliation{Campion College and the Department of Physics, University of Regina, Regina, SK S4S 0A2, Canada}

\author[0000-0001-7244-6069]{Ying-Tung Chen (\begin{CJK*}{UTF8}{bkai}陳英同\end{CJK*})}
\affiliation{Institute of Astronomy and Astrophysics, Academia Sinica; 11F of AS/NTU Astronomy-Mathematics Building, No. 1 Roosevelt Rd., Sec. 4, Taipei 10617, Taiwan}

\author[0000-0002-3376-7297]{Arcelia Hermosillo Ruiz}
\affiliation{Astronomy and Astrophysics, University of California Santa Cruz}

\author[0009-0003-4483-4314]{Cameron Semenchuck}
\affiliation{Campion College and the Department of Physics, University of Regina, Regina, SK S4S 0A2, Canada}
\affiliation{Department of Physics and Astronomy, University of Victoria, Victoria, BC, Canada}

\author[0009-0004-7149-5212]{Cameron Collyer}
\affiliation{Planetary Sciences Group, Department of Physics, University of Central Florida, 4111 Libra Drive, Orlando, FL 32816, USA}
\affiliation{Center for Astrophysics | Harvard \& Smithsonian; 60 Garden Street, Cambridge, MA, 02138, USA}

\author[0000-0001-7032-5255]{J.~J. Kavelaars}
\affiliation{Herzberg Astronomy and Astrophysics Research Centre, National Research Council of Canada, 5071 West Saanich Rd, Victoria, BC V9E 2E7, Canada}
\affiliation{Department of Physics and Astronomy, University of Victoria, Victoria, BC, Canada}

\author[0000-0002-9179-8323]{Lowell Peltier}
\affiliation{Department of Physics and Astronomy, University of Victoria, Victoria, BC, Canada}

\begin{abstract}

The Large inclination Distant Objects (\lido) survey has discovered the first securely classified object in the 10:1 mean motion resonance of Neptune.
This object, 2020 VN$_{40}$, is short-term stable in the 10:1 resonance, but not stable on Gyr timescales.
2020 VN$_{40}$ is likely part of the scattering sticking population, and temporarily resides in the 10:1 resonance at $\sim139.5$~au.
This discovery confirms that this distant resonance is populated, as a single detection is likely to be indicative of a large population that is difficult to detect due to observational biases.
This object has an inclination of 33.4$\degr$, and n-body integrations of orbital clones of 2020 VN$_{40}$ have revealed some unexpected evolutions.
While clones of 2020 VN$_{40}$ show resonant libration around the expected resonance centers of approximately 90$\degr$, 180$\degr$, and 270\degr, for a restricted range of inclination and eccentricity values some clones librate around a resonant argument of 0$\degr$. 
As this occurs for the slightly lower-eccentricity portions of the evolution, this behavior can also be quite stable.
Our initial exploration suggests that this libration around a center of 0$\degr$ is a generic effect for highly inclined objects in $n:1$ resonances because the nature of their resonant interaction with Neptune becomes a strong function of their argument of pericenter, $\omega$.
At large inclination, the resonant islands shift as $\omega$ precesses, switching the center of symmetric libration to 0$\degr$ for $\omega=90\degr$ and $\omega=270\degr$.
2020 VN$_{40}$ provides interesting insight into the evolution of the large-inclination resonators, which become more common at increasing semi-major axis.

\end{abstract}

\section{Introduction}\label{sec:intro}

The distant mean motion resonances (MMR) of Neptune are expected to be populated with trans-Neptunian objects (TNOs) based on both simulations and previous surveys.
Simulations of giant planet migration \citep[e.g.][]{Kaib2016, pikelawler2017, nesvorny2015b} show an extended resonant TNO population with semi-major axes $>55$~au, and large eccentricities ($e\gtrsim0.2$).
The populations of larger-pericenter ($q>40$~au) resonant TNOs also preferentially have larger inclinations $i$ \citep{pikelawler2017}.
The distant resonances are expected to be populated by a combination of scattering sticking \citep[e.g][]{Lykawka:2007} and scattering capture during the era of giant planet migration \citep[e.g.][]{Kaib2016}.
The distant $n$:1 and $n$:2 resonances are particularly efficient at resonant sticking, and are expected to have large populations of dynamically excited TNOs, with up to 40\% of the scattering objects out to the 6:1 ($\sim100$~au) being transiently stuck in resonance at any given time \citep{Yu2018}.
For high-eccentricity orbits with perihelion distances near Neptune, the efficiency of scattering sticking drops off at semi-major axes beyond $\sim250$~au \citep{Lykawka:2007,Graham:2024}.
However, Neptune's MMRs remain strong for more weakly scattering/diffusing TNOs with larger perihelion distances out to semi-major axes of $\sim500$~au \citep{volk2022}.
Determining whether individual TNOs discovered in these distant resonances are more likely to be transiently sticking or long-term stable is critical to understanding how these distant resonances were populated.

The orbital distribution of the distant resonators has not yet been well constrained by surveys, but the orbits of the known distant resonant TNOs have eccentricities and inclinations similar to the scattering and detached TNO populations.
The larger-$e$ values of the observed TNOs could be a selection effect due to observing biases, as the lower-$q$ TNOs are easier to detect, and at $a\gtrsim60$ most surveys cannot conclusively rule out the presence of low-$e$ TNOs \citep[see discussion in, e.g.,][]{Gladman2021}.
Ecliptic surveys such as the Canada-France Ecliptic Plane Survey \citep[CFEPS,][]{petit11} are biased towards discovering low-$i$ objects, but discovered the first known 5:1 resonator which had an inclination of 20.9$\degr$; the subsequent additional 5:1 discoveries in the off-ecliptic CFEPS High Latitude component \citep[HiLat,][]{hilat} were all also found to have larger inclinations \citep{gladman2012,pike2015}.
The 5:1 resonator orbits and related modeling work \citep{pike2015} are consistent with a population of distant resonators which is significantly dynamically excited compared to the classical belt.
The Outer Solar System Origins Survey \citep[OSSOS,][]{bannister2018} discovered 34 distant resonators (which we define as objects in resonances beyond the 2:1 but excluding the 5:2), including two securely classified 9:1 resonators and insecure resonators in the 21:2 and 27:4 MMRs.
Like the earlier 5:1 discoveries, the two OSSOS 9:1 resonators also have larger inclination, and inclinations of 25-40$\degr$ were used in a parametric model of the inclination distribution of this population \citep{Volk2018}.
The confirmed 8:1 and candidate 10:1 resonator in \citep{bernardinelli2022} also have large inclinations of 28$\degr$ and 44$\degr$.
The 22 securely classified distant resonators in OSSOS and related surveys \citep[OSSOS++,][]{petit11,hilat,alexandersen2016,bannister2018} were modeled using similar parametric models, and inclination widths of 14.5-25$\degr$ and non-circular orbits with $q$ near Neptune \citep{Crompvoets2022}.
The larger inclinations of these distant resonators means they are more detectable in off-ecliptic surveys.

When determining the detectability of a particular TNO population in a survey, we often use parametric models of the population \citep[see, e.g.][]{lawlerFASS}.
For resonant TNO populations, this includes not just the $a$, $e$, and $i$ values resonators could have, but also requires constraints on the angular positions where the resonators may be on sky relative to Neptune \citep[e.g.][]{gladman2012,lawler13}.
In particular, resonant TNO populations come to perihelion at a range of angles relative to Neptune that are controlled by the dynamics of the resonance they are in; i.e., their perihelia locations (where they are brightest and most observable) are not uniformly distributed on the sky but instead librate around fixed locations relative to Neptune \citep[for a more detailed discussion see][]{Gladman2021}.
We expect the perihelion directions of a population of $n$:1 resonators to be distributed amongst three libration `islands', the two so-called `asymmetric' islands approximately 90$\degr$ ahead of and 90$\degr$ behind Neptune and the `symmetric' island centered at 180$\degr$ from Neptune. 
Objects in the asymmetric islands have relatively small libration amplitudes ($\lesssim100\degr$) around an eccentricity-dependent libration center (see, e.g., \citealt{Nesvorny:2001} for detailed descriptions of the 2:1 asymmetric islands), while the symmetric island has a fixed libration center and large libration amplitudes ($\sim300-355\degr$); these three libration modes are similar to the familiar tadpole and horseshoe co-orbital libration islands of the 1:1 resonance.
The relative populations within these libration islands is usually treated as another free parameter in parametric modeling \citep[as done in, e.g.,][]{chen2019_twotino}.

Here we find that the simple $n$:1 resonator model described above is not adequate to describe the resonant behavior of moderate-to-large inclination TNOs with perihelia far out of Neptune's orbital plane. 
Numerical integrations of some clones of the observed 10:1 TNO reported in this paper showed an unexpected mode of libration, one where the perihelion location experiences large-amplitude libration around $0\degr$ relative to Neptune.
The potential for libration around $0\degr$ in Neptune's external MMRs has been identified for high-eccentricity, deeply Neptune-crossing orbits \citep{Malhotra:2018,Lan:2019}; for perihelion distances inside Neptune's orbit in the coplanar (zero-inclination) case studied in those works, libration around zero allows the particle to avoid Neptune at perihelion.
For the 10:1 object in this paper, the libration around $0\degr$ is quite different as it occurs for lower-eccentricity orbits with perihelion distances well outside Neptune's orbit and at significant inclination. 
Objects librating in this novel manner would not be accounted for in the asymmetric and symmetric islands of a typical $n$:1 resonance model.
They will also have a different on-sky distribution than predicted by those models, with important implications for interpreting the results of surveys that are sensitive to higher-inclination resonant objects.

In this work we discuss the discovery of 2020 VN$_{40}$ in the \lido survey (Section \ref{sec:discovery}), which used two off-ecliptic blocks to preferentially discover mid- to large- inclination TNOs.
Using the \lido astrometry and some targeted follow-up, we calculate the orbit of the 10:1 resonator 2020 VN$_{40}$ (Section \ref{sec:orbit}).
This includes an explanation of the tracking and orbit fitting, as well as investigations of the long-term evolution of clones of the object.
We investigate the unexpected libration around $0\degr$ in Section \ref{lib0}.
The results are summarized and discussed in Section \ref{sec:discuss}.

\section{The \lido Survey: A search for distant objects with mid to large inclinations}
\label{sec:discovery}

A comprehensive survey description is provided in \citet{Alexandersen2023,alexandersen2025}, and a brief overview is given here.
\lido utilizes a similar discovery and tracking strategy to the Outer Solar System Origins survey \citep[OSSOS,][]{bannister2018}, with minor modifications resulting in a fainter limiting magnitude than OSSOS. 
The survey was conducted on the Canada-France-Hawaii Telescope (CFHT) on Maunakea\footnote{CFHT program numbers 20AC02, 20BC019, 21AC013, 21BC003, 22AC021, 22BC018, 23AC016 (Canada, PI: S.~Lawler) and 20AT02, 20BT005, 21AT004, 21BT003, 22AT011 (Taiwan, PI: Y.~T.\ Chen)}.
Images were acquired using MegaCam and the wide band filter ($gri$) with 320 second exposures.
The two survey blocks are comprised of 18 contiguous MegaCam pointings (2021A) and 16 contiguous MegaCam pointings (2020B).
Object tracking images were acquired using CFHT MegaCam in 2020A through 2023A.
The images were processed in the CFHT pipeline, and the World Coordinate System (WCS) and photometric zero points were determined with high accuracy using MegaPipe \citep{Gwyn2008}.
The discovery images were acquired with a nearly identical survey strategy to the OSSOS survey, with a triplet of three images on one night.
During the discovery year, blind tracking was conducted, on at least one additional night within the discovery dark run, and additional images in other months, with the fields shifted based on a typical TNO rate of motion to maximize recovery.
Discovery, characterization, and tracking were done using the OSSOS Moving Object Pipeline (MOP) and object tracking tools \citep{bannister2018}.
After the discovery semesters, tracking was targeted based on ephemerides calculated from the survey's astrometry.
The survey discovered and tracked 141 objects.
The survey characterization limit was carefully quantified for the different pointings within the two survey blocks, and ranging between $m_{gri}$=25.0--25.6.
16 tracked discoveries were fainter than the characterization limit or outside the rate of motion limit of the survey, so the characterized sample in \lido is 125 objects and includes at least 56 objects in mean motion resonances with Neptune.

\section{Orbital Characterization of 2020 VN$_{40}$}
\label{sec:orbit}

The \lido survey was designed to discover TNOs and track those discoveries over two additional years.
This orbital arc length was selected to facilitate the classification of resonant TNOs with semi-major axes near the classical belt, especially in the 3:2 resonance with Neptune.
The 10:1 resonator discovered in the survey, 2020 VN$_{40}$, was tracked from August 2020 until October 2022 as part of the main \lido Survey. 
At the end of the planned \lido Survey, the orbit of 2020 VN$_{40}$ and a few other large-$a$ objects had not quite met the target orbital uncertainty ($\delta$$a$/$a$=0.001--0.0001), in spite of 2020 VN$_{40}$ having an arc length of 806 days.
The \lido observational arcs are still short compared to the orbital periods of these large-$a$ objects, so their uncertainties do not converge as quickly as for the Plutinos which were the design focus of the survey. 
Additionally, dynamical classification of large-$a$ objects requires increased fractional accuracy compared to Plutinos because more distant resonances have smaller fractional widths than closer-in resonances \citep[e.g.][]{Lan:2019}.
Using only astrometry from the the main \lido survey, the ProjectPluto\footnote{\url{https://www.projectpluto.com/}} find\_orb \citep{Gray2022} orbit for 2020 VN$_{40}$ was a semi-major axis of $a=139.99\pm0.19$~au, or a relative uncertainty $\delta$$a/a=$0.0014.
This uncertainty did not fully lie within the phase space of the 10:1 resonance, so additional data beyond the planned \lido Survey was required for this object.

\subsection{Additional Astrometric Data}
We acquired additional tracking in January and February of 2024 using the Gemini Multi-Object Spectrograph \citep[GMOS,][]{hook2004} on Gemini North on Maunakea through program GN-2023B-DD-110 (PI R. Pike) to further refine the orbit of 2020 VN$_{40}$.
The GMOS images were processed using the Gemini \textit{pyraf} data reduction package to do the flat and bias correction and combine the chips into a mosaic.
The plate solution was refined using astrometry.net\footnote{\url{https://astrometry.net/}} software \citep{Lang2010} and the Pan-STARRS PS1PV3 \citep{Schlafly2012, Tonry2012, Magnier2013}.
The Pan-STARRS DR2 PS1PV3 uses Gaia as an absolute reference, cross-matching their common stars to correct systematic errors and refine the Pan-STARRS astrometric solution.
This effectively anchors Pan-STARRS PS1PV3 to the highly precise Gaia frame and makes it easier to extrapolate an accurate world coordinate system to fainter stars than the Gaia sources \citep{lubow2021}.
This provides a sufficiently large catalog of Gaia-based source positions that are unsaturated in our images for a robust plate solution.
The position of the TNO in each image was then measured using \textit{SourceExtractor} \citep{Bertin1996}.
With the increased orbital refinement from 2024, we were able to identify 2020 VN$_{40}$ in five archival Subaru Hyper Suprime-Cam images from two nights in September 2017, using a Solar System Object Image Search  \citep[SSOIS,][]{Gwyn2012} to identify possible observations.
The Subaru images from September 2017 were downloaded and processed with HSCPipe 8.3 \citep{Bosch2018}, including a plate solution refinement from Pan-STARRS PS1PV3.
The predicted position of the TNO based on the ephemeris was accurate within a few arcseconds, and the transient nature of the source was confirmed by comparing the multiple September 2017 images.
The position of 2020 VN$_{40}$ was measured using \textit{SourceExtractor}.
This additional astrometry brought the arc length to 2,323 days, and an orbital uncertainty from ProjectPluto of $\delta$$a/a=$0.00032.
The barycentric orbit of  2020 VN$_{40}$ using all of the available astrometry is $a$=139.95$\pm$0.05~au, $e$=0.72661$\pm$0.00009, $i$=33.4070$\degr$$\pm$0.0001$\degr$, $\Omega$=197.3069$\degr$$\pm$0.0002$\degr$, $\omega$=262.892$\degr$$\pm$0.004$\degr$, $M$=355.225$\degr$$\pm$0.002$\degr$, which is a pericenter $q$=38.26~au.
This improved orbit was sufficiently precise to conclusively identify this TNO as resonant in Neptune's 10:1 mean motion resonance.

\subsection{Resonant Classification}
\begin{figure}
    \centering \includegraphics[width=0.8\textwidth]{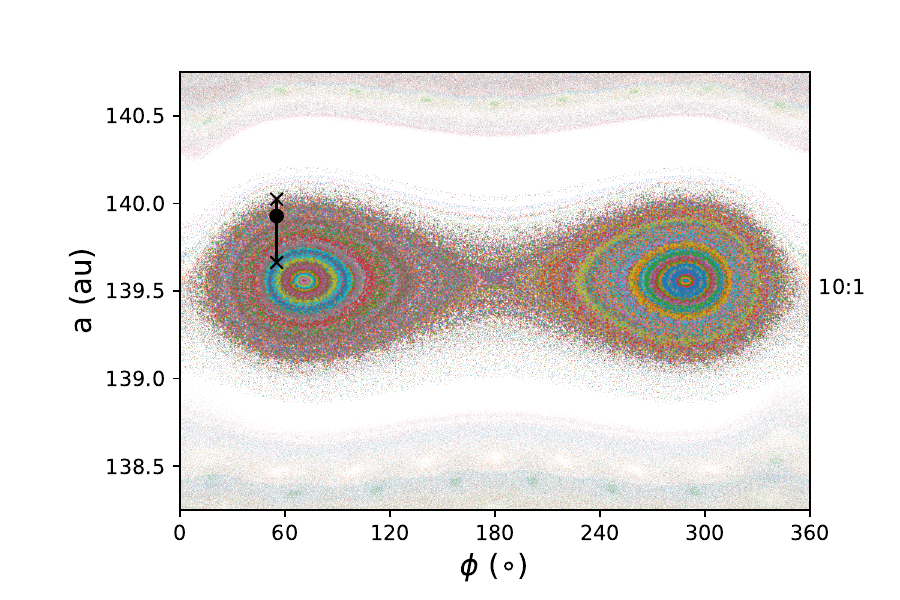}
    \caption{
    A Poincar\'e Map of the semimajor axis vs resonant angle phase space near the \lido 10:1 resonator.
    The best-fit orbit of 2020 VN$_{40}$ is shown as a black dot connected to the 3-sigma high- and low-$a$ orbit fits (black `x's; the uncertainty in $\phi$ is negligible and not shown).
    The colored points show the evolution of test particles with the same perihelion distance, inclination, and argument of pericenter as 2020 VN$_{40}$, but with different initial values of $\phi$ and $a$. 
    Each particle is shown in a different color with 10:1 resonant particles shown with larger, less transparent points; we plot $a$ and $\phi$ for each particle at every perihelion passage over 1500 orbits ($\sim2\times10^5$ years).
    The only significant resonant behavior in the displayed phase space is the 10:1. 
    The closest resonances to the 10:1 (the 41:4, 51:5, 61:6) are well outside the uncertainty range of 2020 VN$_{40}$.
    The uncertainty in 2020 VN$_{40}$'s orbit falls completely within the libration zone of the 10:1 resonance.
    }  
    \label{fig:poincare}
\end{figure}

2020 VN$_{40}$ has a semi-major axis within the boundaries of the 10:1 MMR, and this object appears to be experiencing resonance-sticking in the 10:1.
2020 VN$_{40}$ is both high-$e$ (0.727) and high-$i$ (33.407$\degr$), and is near the survey limiting magnitude, with an apparent magnitude of $m_{gri}$=24.7 and an absolute magnitude of $H_{gri}$=8.32.
We use two methods to understand and characterize the orbit of 2020 VN$_{40}$: the creation of a Poincar\'e map to characterize the orbital phase space near the observed orbit and long-term integration of 200 clones based on the astrometric uncertainty of the specific object.
The Poincar\'e map, shown in Figure~\ref{fig:poincare}, is a way of visualizing resonant libration islands over a range of semimajor axes for fixed values of perihelion distance, inclination, and argument of perihelion. 
Following \cite{volk2022}, we integrated the orbits of several thousand test particles with 2020 VN$_{40}$'s perihelion distance, inclination, and argument of perihelion, but with semi-major axes and initial resonant angles spanning the range seen in Figure~\ref{fig:poincare}, under the gravitational influence of the Sun and all four giant planets for $2\time10^5$ years. 
We record the semi-major axis and resonant angle of each particle every time it passes through perihelion to generate Figure~\ref{fig:poincare}.
This reveals the three expected libration islands of the 10:1, the two asymmetric islands around $\sim90\degr$ and $\sim270\degr$ and the symmetric island centered around $180\degr$. 
We discuss the dynamical structure of the 10:1 in more detail in Section~\ref{ss:dynamics}, but Figure~\ref{fig:poincare} shows that 2020 VN$_{40}$'s current semi-major axis uncertainty range completely overlaps the well-defined libration islands of the 10:1.
This, along with the clone integrations discussed below, indicate that 2020 VN$_{40}$ is currently securely librating in Neptune's 10:1 MMR, with the integrations providing insight into the long-term behavior of the TNO and the 10:1 resonance at large inclinations.

\begin{figure}
    \centering \includegraphics[width=0.65\textwidth]{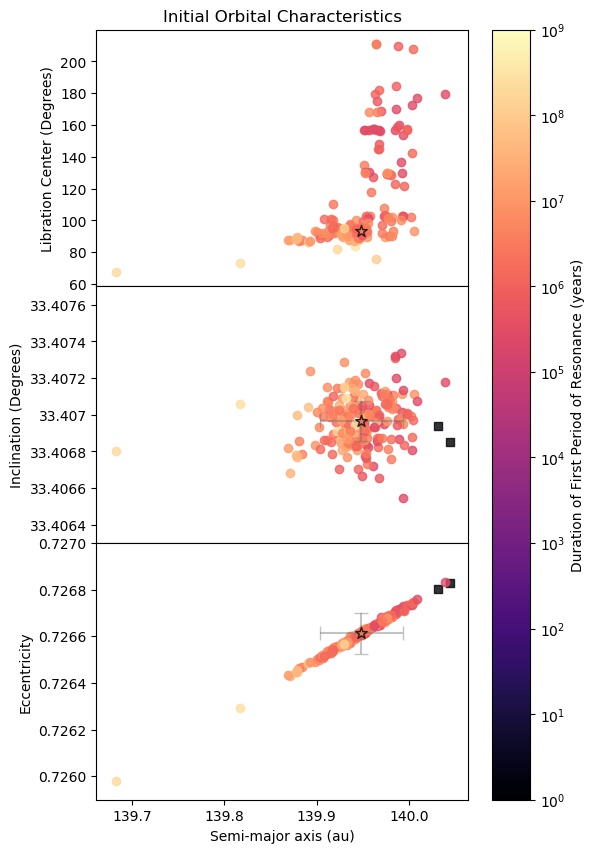}   
    \caption{The initial semi-major axis, eccentricity, inclination, and libration amplitude of the 200 clones of 2020 VN$_{40}$ and the unfuzzed nominal orbit (star).  The find\_orb uncertainties on the nominal orbit are shown in gray.  The color of the points is the amount of time spent in a single resonant island starting at time 0, or the first period of resonance.  The libration center is calculated only over the first period of resonance within the first resonant island; further into the simulations the libration center can include centers around 270$\degr$ and 0$\degr$.  The clones are created by fuzzing the astrometry based on expected uncertainty, and repeating the orbital fitting process. Two clones are not resonant at time step 0, and are indicated by black squares. There are definite trends including a dependence between $a$ and stability and $a$ and libration center.  The larger-$a$ points are less stable and include symmetric librators in addition to asymmetric librators.}  
    \label{fig:fuzz_orbits}
\end{figure}

To fully understand the range of possible orbital evolutions for 2020 VN$_{40}$, we generate orbits that are consistent with the measured astrometry and sample the orbital uncertainty in a representative way.
To do this, we resample (or fuzz) the measured astrometry around the nominal position measurements.
For each point, we randomly generated a new astrometric measurement from a Gaussian distribution centered on the actual measurement with the width of the distribution being the 1$\sigma$ uncertainty.
While early TNO observations sometimes suffered from systematic astrometric errors (see, e.g., discussion in \citealt{gladman2008}), astrometric measurements in the Gaia era appear to be  nearly Gaussian, meaning the orbit-fits derived from this assumption provide a reasonable measurement of the orbital uncertainties \citep[e.g.][]{volk2016}.
The 1$\sigma$ orbital uncertainty as computed by find\_orb is also shown in Figure \ref{fig:fuzz_orbits} for comparison with our orbit distribution. 
We tested generating up to 1000 clones, and found that using more than 200 clones had minimal effect on the resulting distribution.
The 200 clones sample approximately the 3$\sigma$ uncertainty on the object's orbit.
For 2020 VN$_{40}$, all of the astrometry available for this object was acquired from telescopes on Maunakea between 2017-2024 and was carefully calibrated by our team.
We selected the typical uncertainty value from the Minor Planet Center (MPC) \citep{Dziadurainprep} which matched the observatory code for Maunakea and the utilization of a Gaia catalog, 0.1$\arcsec$.
ProjectPluto find\_orb was used to fit the nominal orbit (without uncertainty fuzzing) and 200 clones based on fuzzed astrometry, see Figure \ref{fig:fuzz_orbits}.
Fuzzing and re-fitting the orbit retains the interdependence of some orbital elements, such as $a$ and $e$, and can be used to generate a representative distribution of clones in orbital element space.
The cloned orbits are all consistent with the object's astrometry and are shown in Figure \ref{fig:fuzz_orbits}.
These orbits provide a statistical overview of the likely behavior of the TNO over time.

\begin{figure}[h]
    \centering \includegraphics[width=1.1\textwidth]{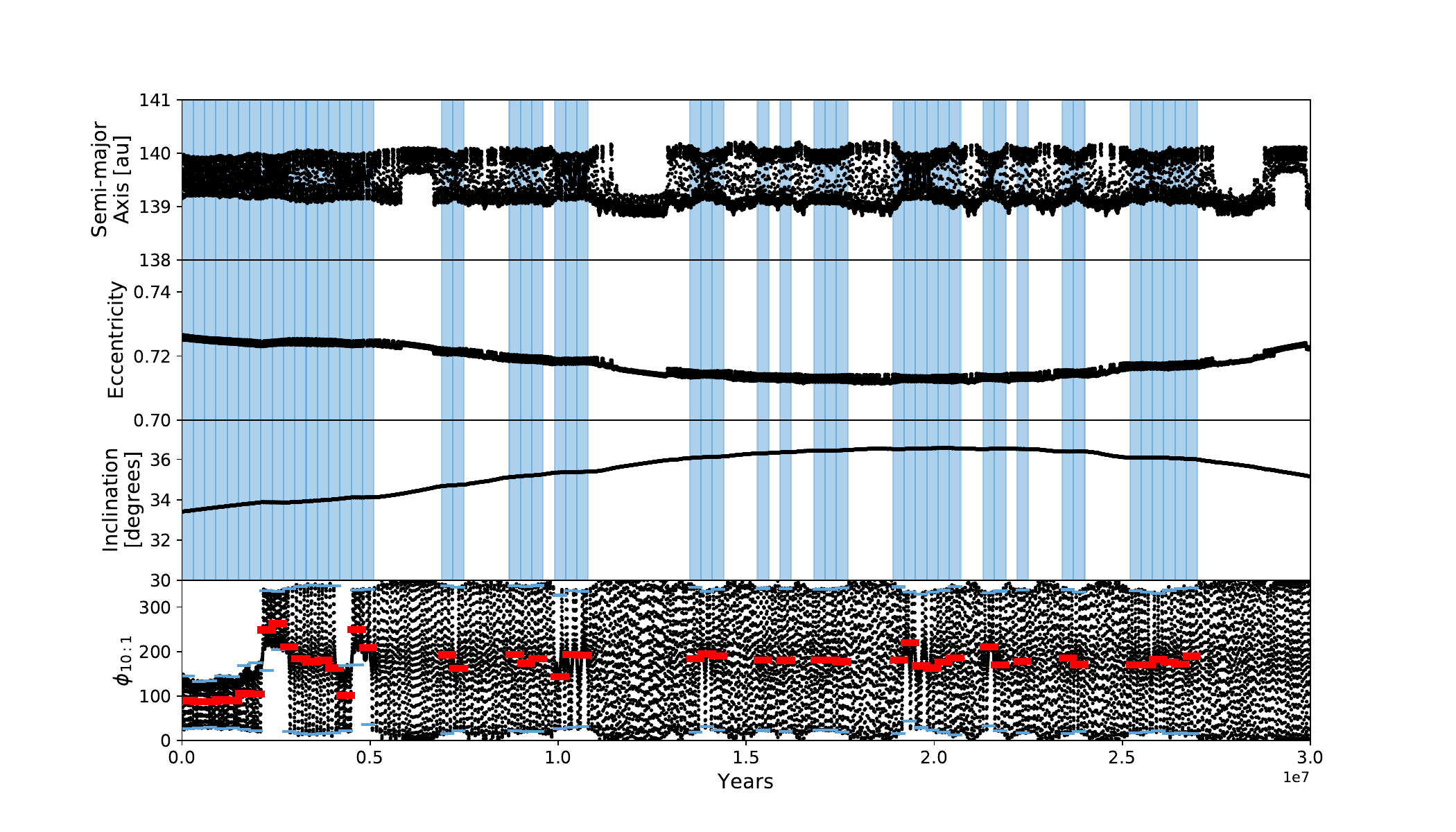}
    \caption{Nominal clone behavior over the first 30 Myrs.  The semi-major axis, eccentricity, inclination, and resonant angle are shown over time.  The blue regions indicate the windows during which the classification code determined that the particle was resonant.  In the $\phi$ panel, the red lines show the libration center and the blue lines show the resonance bounds which are used to determine the libration amplitude.  The nominal clone begins in the leading asymmetric island, which is the most common starting behavior for the clones, and spends 50\% of the first 30~Myrs in resonance.
    }  
    \label{fig:nominal}
\end{figure}

The nominal orbit and 200 clones were integrated for 30~Myr and 1~Gyr to determine their orbital classifications and long-term behavior.
We used the WHFast integrator \citep{rein2015} through REBOUND \citep{rein2012}, with the four giant planets and the Sun included as perturbers (the mass of the terrestrial planets was added to the Sun).
The REBOUND simulations used a time step of 0.5~year and output frequency of 2000 years.
This frequent sampling was required to clearly identify individual librations of the resonant angle $\phi$.
The REBOUND output was analyzed using a classification code, which divided the output into 300,000 year windows.
We created an updated version of the resonant classification code from \citet{lawler2019}, and heavily tailored the criteria to the 10:1 resonance to increase the accuracy of resonant identification for this specific object.
The classification code uses a windowing and grid methodology similar to previous works \citep[e.g.][]{lawler2019, khain2020}.
In each window, the classification code tested whether the average semi-major axis was within 0.06 au of the resonance center.
This width was determined to be appropriate for this object based on the results from multiple iterations of the classification script with different widths. 
The classification script then calculated the resonant angle
\begin{equation}
    \label{eq:phi}
    \phi = 10\lambda_{\rm TNO} - \lambda_{\rm Neptune} - 9\varpi_{\rm TNO},
\end{equation}
where $\lambda$ is the mean longitude of the TNO and Neptune and $\varpi_{\rm TNO}=\omega + \Omega$ is the TNO's longitude of perihelion, the sum of the argument of perihelion ($\omega$) and the longitude of ascending node ($\Omega$).
The classification script then divided $\phi$ values within each time window into blocks of 5$\degr$ in $\phi$.
Due to the high sampling of the resonant angle, libration in $\phi$ could be easily identified by searching for windows containing empty blocks in $\phi$.
A window was flagged as resonant if any of the following criteria were true: (1) the 0-5$\degr$ and 355-360$\degr$ blocks were empty (indicating symmetric or asymmetric libration), (2) 40\% of the blocks were empty (indicating asymmetric libration), or (3) the two blocks at 175-185$\degr$ were empty (indicating libration around $\phi=0$; see Section \ref{lib0} for details).
If the window was identified as resonant, the classification code then determined the mean value of $\phi$ within the window to be the resonant center, and the upper and lower bounds of the libration were determined by sorting the $\phi$ values and identifying the $3\sigma$ upper and lower bounds.
(Note that if the resonant center was near zero, the data was appropriately shifted in order to calculate accurate values for the upper and lower bounds and the mean.)
The upper and lower $\phi$ values were then used to compute a libration amplitude.
The classification of the nominal clone is shown in Figure \ref{fig:nominal}, including the window width, resonant center, and upper and lower $\phi$ bounds that are used to compute the libration amplitude.
The results of this code were plotted and random human checks were done to ensure that the resulting classifications and characteristics were reasonable.

The possibility of mixed inclination-eccentricity resonances were also briefly explored.
The same classification methodology was used, with revised $\phi$ equations as appropriate \citep{murray_dermott2000}.
Only a few short periods of mixed mode resonance were identified over the Gyr simulations of the 201 clones, so this behavior was found to not be representative of the likely current or future orbit of the real TNO.

\begin{figure}[h]
    \centering \includegraphics[width=1.\textwidth]{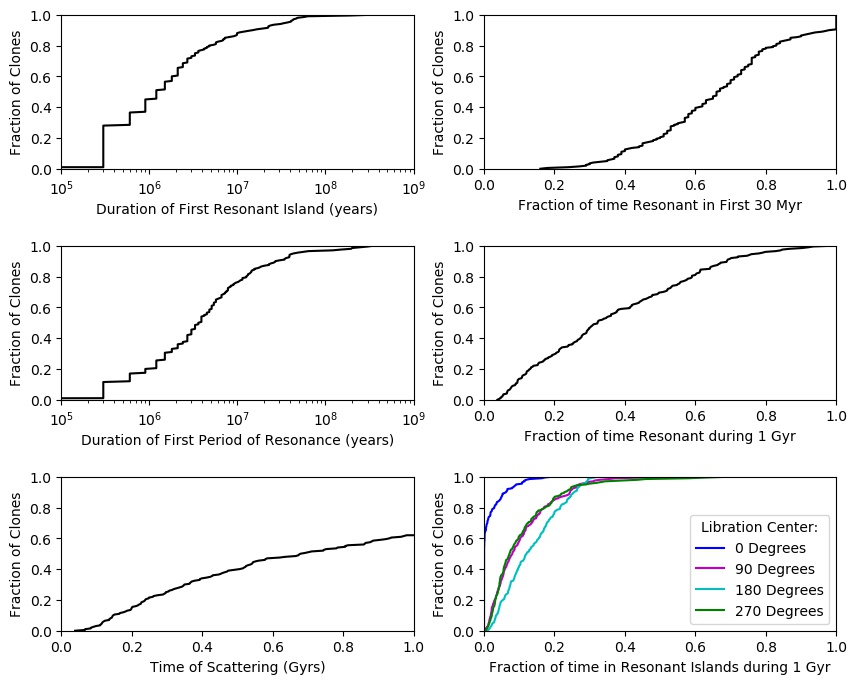}
    \caption{The fraction of time clones of 2020 VN$_{40}$ spend in various orbital states.  The duration spent in the first period of resonance can include occupancy in multiple resonant islands, as long as the test particle's $\phi$ does not circulate between these modes.  The first resonant island includes only the first mode. In the first 30 Myr, 50\% of the particles are resonant 66\% of the time, and in 1 Gyr, 50\% of the particles are resonant 32\% of the time. The particles begin to scatter at 39~Myrs, and 62\% have scattered within 1 Gyr, indicating that the object is likely not stable on Gyr timescales.  The amount of time clones spend in each of the leading and trailing island are comparable, while more time is spent in the symmetric island. The libration around 0$\degr$ is detected for 51\% of clones, and while it is not as common as the other libration modes it does represent a significant fraction of the long-term behavior.}  
    \label{fig:times}
\end{figure}

In order to classify 2020 VN$_{40}$, we examine the first 30~Myrs as well as the full 1 Gyr simulations.
In the first 300,000 year window of the simulation, 199/201 particles (including the nominal clone) are classified as resonant.
The two non-resonant clones at the start of the simulation are two of the three largest-$a$ clones, and are significantly less likely to indicate the behavior of the real TNO.
Additionally, while they are not resonant at the start of the simulation, they both show multiple periods of resonance during the simulation, and appear to be repeatedly sticking in the 10:1.
(The minimum fraction of time that any particle spends in resonance during the first 30~Myrs is 16\%.)
The particles are shown in Figure \ref{fig:fuzz_orbits}, colored to indicate the duration of their first period of resonance starting from time step 0.
The first period of resonance is the sum of the time in all resonant windows before the classification code determines that the particle is not resonant.
This can include resonance in multiple different islands, as seen in the nominal clone in Figure \ref{fig:nominal}, where the particle's period of first resonance includes occupying the leading island, the trailing island, and the symmetric island.
Several other metrics were computed to evaluate the evolutionary behavior of the particles.
This included the duration of first resonance in a specific island (for the nominal, this would be the duration of resonance in the leading island), the fraction of time the particle was resonant in the first 30 Myr, and the fraction of time the particle was resonant in the 1 Gyr simulation.
The fraction of time in resonance was also split by resonant island, identified using 90$\degr$ bins for the libration center (average $\phi$) value in the window.
This split the fraction of time resonant into the leading, trailing, and symmetric islands as well as libration around $\phi=0\degr$ (see Section \ref{lib0} for details).
Additionally, the time of scattering was identified as the first time that the particle's average semi-major axis in a window was more than 2 au from the center of the 10:1 resonance.
These bulk properties of the clones, shown in Figure \ref{fig:times}, provide a useful framework for understanding the expected behavior of the real object 2020 VN$_{40}$.

\begin{figure}[h]
    \centering \includegraphics[width=.55\textwidth]{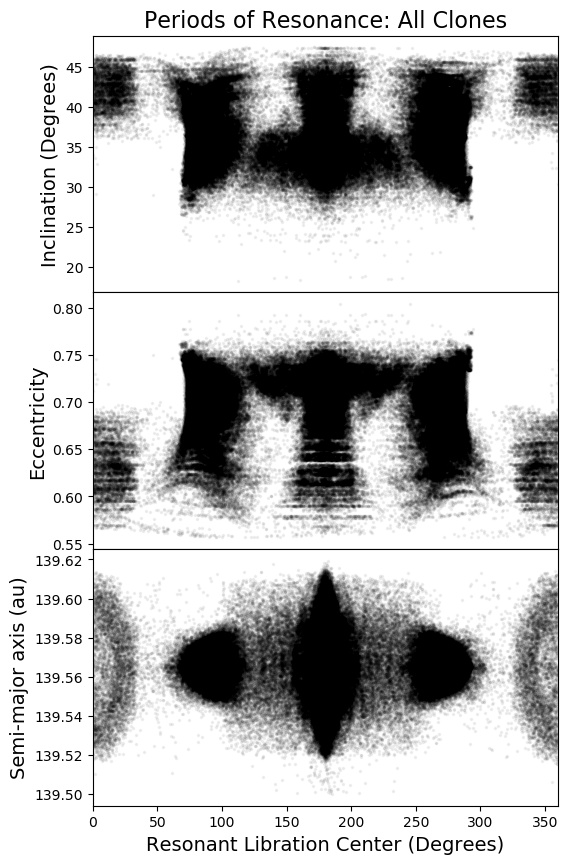}
    \caption{The resonant behavior of the clones of 2020 VN$_{40}$.  Each point is the average value from a 300,000 year window in the resonant analysis.  Each window where the particle is classified as resonant is plotted for all 201 clones. The libration around  $\phi= 90\degr, 180\degr$, and $270\degr$ are clearly the dominant modes, however, a not insignificant fraction of windows show the libration around  $\phi=0\degr$.  The different resonant islands also have different dependencies on $a$, $e$, and $i$, with $\phi=0\degr$ having a more limited range of $e$ and $i$.  The diffuse points between the darker clumps of the three dominant libration centers in the $a$ panel represent particles transitioning between different resonant islands, where determining the resonant libration center with a simple average is less accurate.  The apparent `bands' in the $e$ plot are simply an artifact of the sampling; in areas where there are fewer clones, individual clones can become more distinct.}  
    \label{fig:res_center}
\end{figure}

The clones of 2020 VN$_{40}$ show a variety of resonant behavior over the 1 Gyr simulations.
The clones occupy multiple resonant islands while they are in and near the 10:1 resonance, and approximately 60\% of the clones scatter by the end of the 1 Gyr simulations.
As another way to visualize the time spent in different resonant configurations, we examined the behaviour of $a$, $e$, $i$ compared to the resonance center in each window.
Figure \ref{fig:res_center} shows all 201 clones and includes one point for each window where a particle was identified as resonant. 
The average $a$, $e$, $i$, and resonant center are computed over that window.
The high-density regions indicate that the clones spend more time in those resonant configurations.
There is a clear interdependence between $e$, $i$, and resonant center.
The particles show libration around $\phi=0\degr, 90\degr, 180\degr$ and $270\degr$.
As discussed in Section~\ref{sec:intro}, libration around $\phi=0\degr$ was not expected. We explore this novel libration mode in the next section. 

\section{The Dynamics of Libration around $\phi=0\degr$}
\label{lib0}
The unexpected libration around $\phi=0\degr$ was immediately apparent in approximately $30$\% of our 1~Gyr simulations, and a further $\sim20$\% showed at least some libration around $\phi=0\degr$.
We did additional investigations to understand the prevalence of this "eyehole" libration behavior for clones of this object and the $n$:1 resonances more generally.

\subsection{Eyehole Libration in the Simulations}
\begin{figure}[!h]
    \centering \includegraphics[width=1.\textwidth]
    {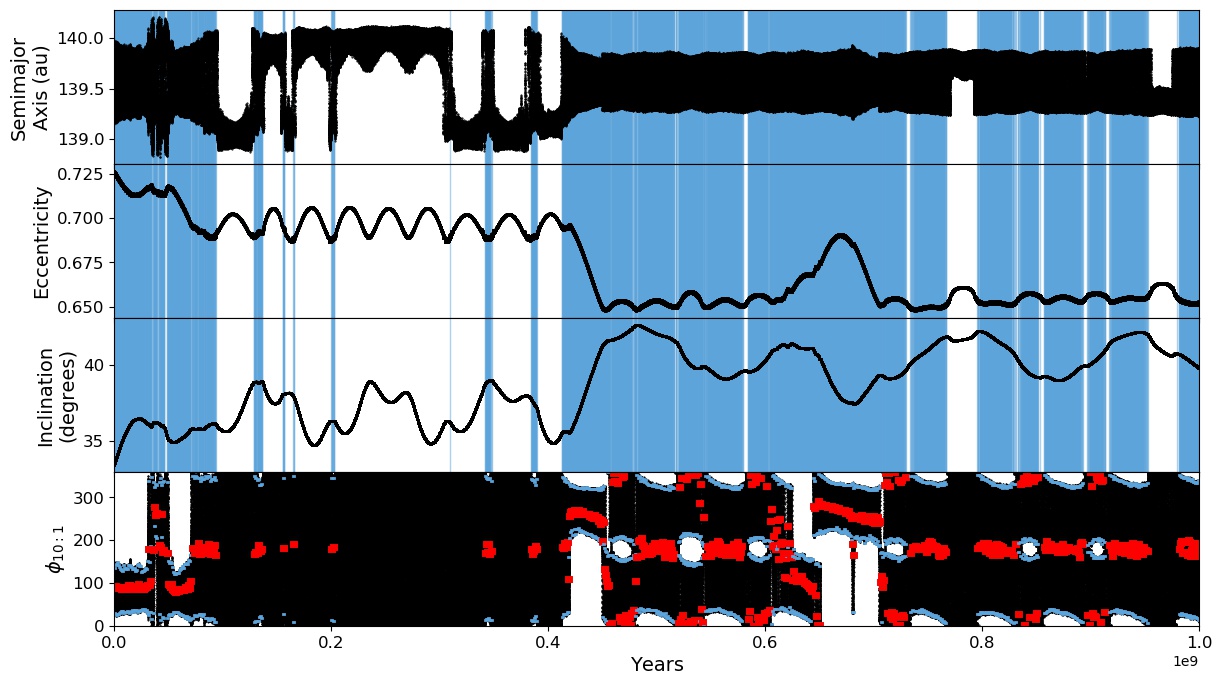}
    \caption{The evolution of one of the clones of 2020 VN$_{40}$ that demonstrates the "eyehole" behavior, particularly between 0.46-0.64 Gyr.  The evolution of this clone is typical for particles showing resonance around $\phi=0\degr$.  This includes transitions between other resonant islands and $\phi=0\degr$, as well as the evolution to higher-$i$ and lower-$e$ before this behavior begins.  As in Figure \ref{fig:nominal}, the periods of resonance are marked in blue and the resonance center and resonance bounds are marked. Each libration of $\phi$ is sampled $>50$ times.}  
    \label{fig:part41}
\end{figure}

We significantly increased our output frequency for the REBOUND simulations to ensure that we were properly interpreting the simulation output and the unexpected libration was not due to a coincidence of a specific output timing.
Each libration includes at least 50 data points, showing clear oscillations.
The libration around $\phi=0\degr$ for one of our test particles is shown in Figure \ref{fig:part41} and the upper panel of Figure \ref{fig:hamiltonian}.
The evolution of this particle is representative of this behavior as seen generally in our sample, including the apparent `eyehole' behavior as the particle repeatedly transitions between oscillations around $\phi=0\degr$ and $\phi=180\degr$.
Transitions between the asymmetric islands and $\phi=0\degr$ are also observed.

We investigated the prevalence of libration around $\phi=0\degr$ in our clones.
This never occurs at the start of our simulations, because the test particles need to evolve to a slightly lower $e$ and higher $i$ than their initial conditions.
Our preliminary testing of the 10:1 and 2:1 resonances indicates that there are likely $e$-dependent critical $i$-values at which this resonant behavior occurs.
This evolution does occur for a significant portion of our clones (51\%, with 21\% having more than 30~Myr of oscillation in this mode), see the lower right panel of Figure \ref{fig:times}.
When particles do show this libration around $\phi=0\degr$, this period of resonance (transitioning between $\phi=0\degr$ to the other resonant islands and back repeatedly) tends to be more stable than the typical symmetric and asymmetric libration, as seen in Figure \ref{fig:part41}.
This is likely due to the lower eccentricity, which reduces the chances of a stronger, scattering perturbation from Neptune.
In Figure \ref{fig:res_center} the libration around $\phi=0\degr$ clearly correlates with relatively higher inclinations and lower eccentricities.

\subsection{The Dynamics of Libration Around $\phi=0\degr$}\label{ss:dynamics}

\begin{figure}[!h]
    \centering \includegraphics[width=0.8\textwidth]{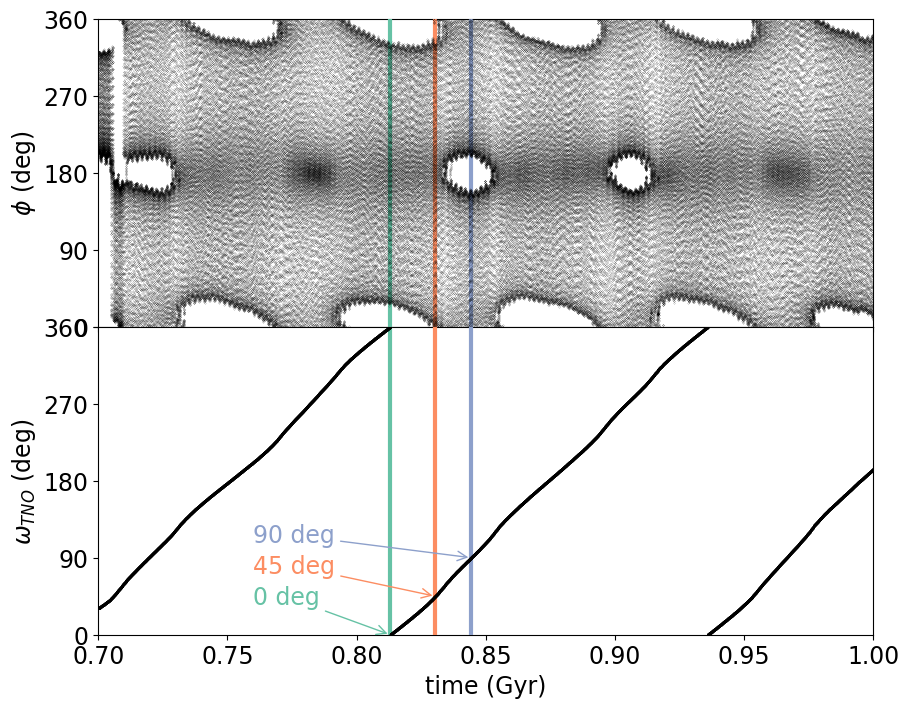}
    \includegraphics[width=1.0\textwidth]{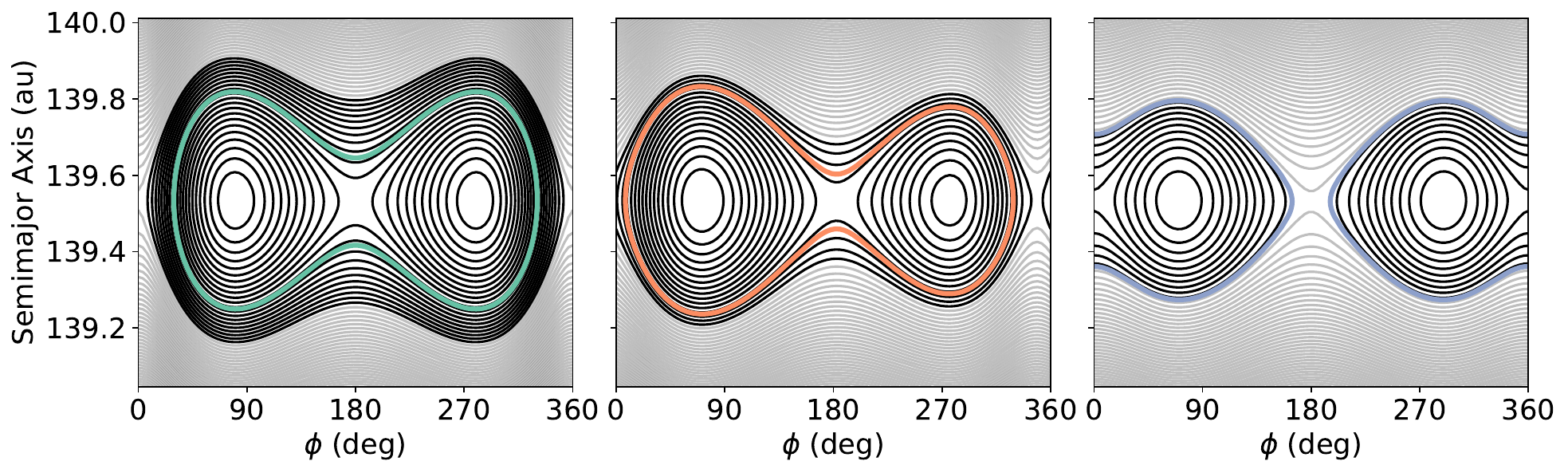}
    \caption{Example section of the clone's evolution in time experiencing ``eyehole" libration, sub-selected from Figure \ref{fig:part41}.  The resonance angle, $\phi$ (\textbf{top panel}), librates about 180$\degr$ when the TNO's argument of pericenter, $\omega_{TNO}$ (\textbf{middle panel}), is near 0$\degr$ (teal line) or 180$\degr$. When $\omega_{TNO}$ is near 90$\degr$ (blue line) or 270$\degr$, $\phi$ librates about $0\degr$.  Transitional behavior (e.g., $\omega_{TNO} = 45\degr$; orange line) results in shifted libration centers for asymmetric libration.  Curves of constant averaged Hamiltonian (\textbf{bottom panel}; see text) illustrate this dynamical behavior in phase space.  Curves are shown for constant $i_{TNO} = 40.5\degr$, a constant canonical momentum $N^\prime$ chosen such that the average value of $e_{TNO}$ is 0.655, and constant $\omega_{TNO} = 0\degr$ (left), $45\degr$ (middle), and $90\degr$ (right).  Each of the three panels displays the same set of Hamiltonian values, and the colored curve in each panel has the same value.  Curves interior to the separatrix (libration) are colored black while those outside the separatrix (circulation) are gray. Note that separatrix values are not consistent across panels meaning that the libration at the constant Hamiltonian value shown in color in each panel is most stable for $\omega_{TNO}=0$.}  
    \label{fig:hamiltonian}
\end{figure}

To explore what drives the change in libration behavior, we analytically model the dynamics of the 10:1 resonance using one of the simulated particles as a reference point.
Figure \ref{fig:hamiltonian} illustrates the dynamical origin of the ``eyehole" behavior frequently seen in the time evolution of $\phi$ for clones of 2020 VN$_{40}$, using the last 0.3 Gyr of the evolution from Figure \ref{fig:part41} as an example.  
Note that the pattern of oscillation between libration about $\phi = 180\degr$ and $\phi = 0\degr$ varies regularly and periodically along with the precession of the test particle's argument of pericenter, $\omega_{TNO}$ (middle panel).  
When $\omega_{TNO} = 0$ (teal vertical line), $\phi$ librates about $180\degr$ as is expected for coplanar 10:1 resonance.  
When $\omega_{TNO} = 90$ (blue vertical line), $\phi$ librates about 0$\degr$, and when $\omega_{TNO} = 45\degr$ (orange vertical line), $\phi$ exhibits transitional behavior.  

Because $\omega_{TNO}$ and inclination both change on a much slower timescale than the libration time, we can verify the dependence of the libration behavior on $\omega$ using a set of Hamiltonian curves, which can be thought of as curves of constant energy in the resonant potential.  
Following the procedure outlined in \citet{murray-clay2005}, we calculate sets of curves of constant Hamiltonian averaged over the synodic period of the resonance (Figure \ref{fig:hamiltonian} bottom panels). 
For the 10:1 resonance, the appropriate canonical coordinates are $\phi$ (Equation \ref{eq:phi}),
synodic angle $\sigma \equiv \lambda_{\rm TNO} - \lambda_{\rm Neptune}$, and $-\Omega_{\rm TNO}$, where $\Omega_{\rm TNO}$ is the particle's longitude of ascending node.  
Their conjugate momenta are $\Gamma^\prime = \Gamma/9$, $N^\prime = \Lambda - (10/9)\Gamma$, and $Z$, where $\Lambda$, $\Gamma$, and $Z$ are the canoncial momenta in Poincar\'e coordinates:
\begin{equation}
    \Lambda=\sqrt{GM_\odot a}, \,\,  \Gamma = \sqrt{GM_\odot a}\left(1 - \sqrt{1-e^2} \right), \,\, Z=\sqrt{GM_\odot a(1-e^2)}(1-\cos{i}).
\end{equation}
In this system of coordinates, we fix $\omega_{TNO}$ and $i_{TNO}$ and average over the full pattern period of the resonance (in this case, 9 orbital synodic periods) so that $N^\prime$ is a constant of the motion, yielding an average Hamiltonian:
\begin{equation}
    H_{\rm avg} = -(GM_\odot)^2/[2(N^\prime + 10\Gamma^\prime)^2] - n_N(\Gamma^\prime + N^\prime) + R_{\rm avg}(\phi, \Gamma^\prime, N^\prime),
\end{equation}
where $n_N$ is the mean orbital angular frequency of Neptune and $R_{\rm avg}$ is the disturbing potential (see \citealt{murray_dermott2000}).  
The term $R_{\rm avg}$ is averaged over the synodic angle, $\sigma$, as described in \citet{murray-clay2005}.  
We choose $i_{TNO} = 40.5\degr$ and $N^\prime$ such that the average $e_{TNO}$ is 0.655 to match representative values in the time window of the integration shown in the top panel of Figure \ref{fig:hamiltonian}.  
Curves of constant  $H_{\rm avg}$ are provided in the bottom row of Figure \ref{fig:hamiltonian} for $\omega_{TNO} = 0\degr$ (left), $\omega_{TNO} = 45\degr$ (middle), and $\omega_{TNO} = 90\degr$ (right), illustrating the transition from symmetric libration about $180\degr$ to symmetric libration about $0\degr$.  
The same range of Hamiltonian values are plotted in each panel with each of the three colored contours having the exact same value of the Hamiltonian. 
The black curves within each panel are interior to the resonant separatrix (i.e., they represent resonant libration of $\phi$) while the gray curves are exterior to the separatrix in each panel (i.e., circulation of $\phi$).  
The teal curve for $\omega_{TNO} = 0\degr$ is farther from the separatrix than the colored curves for other values of $\omega_{TNO}$,  illustrating that symmetric libration about $\phi = 180\degr$ remains strongest for this value of the Hamiltonian. 
Note that the Hamiltonian curves for $\omega_{TNO}=0\degr$ (bottom left panel in Figure~\ref{fig:hamiltonian}) shows the same resonant libration islands seen in the fully numerically generated Poincar\'e map in Figure~\ref{fig:poincare}.
The Hamiltonian calculations consider a simplified system of the Sun, Neptune, and a TNO test particle, so we also generated additional Poincar\'e maps for orbital parameters consistent with those assumed in Figure~\ref{fig:hamiltonian}; the shift in libration center with changing $\omega$ occurs similarly in these full numerical simulations including the Sun, all four giant planets, and test particles.
We also confirmed that Hamiltonian curves for other $n$:1 resonances display the same shifts in the libration island, indicating this is a generic effect for high-inclination orbits.

\begin{figure}[htbp]
    \centering
         \includegraphics[width=0.85\linewidth]{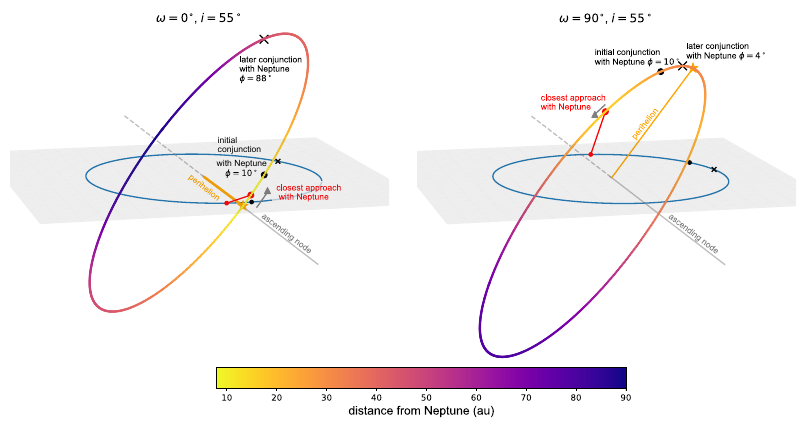} 
    \caption{Visualizations of two highly inclined 2:1 resonant particles completing one orbit (equal to one resonant pattern) with an initial resonant angle $\phi=10^\circ$. The particle orbit is color-coded by its distance from Neptune throughout the resonant pattern and the gray arrow shows the direction of motion. Neptune's orbit is shown in blue. The left panel shows a particle with $\omega=0^\circ$ and the right panel a particle with $\omega=90^\circ$. The black dots show the initial location of conjunction with Neptune, and the connected red dots show the closest approach to Neptune for the orbit with that conjunction. The black Xs indicate the location of conjunction with Neptune ten particle orbits (also ten resonant patterns) later, showing how the perturbations from Neptune move conjunction (and the resonant angle) either farther away from (left panel) or back toward (right panel)  $\phi=0^\circ$.
    }
    \label{f:conjunctions}
\end{figure}

Qualitatively, these different centers of symmetric libration may be understood by considering the orientation of the particle's orbit when it receives gravitational `kicks' near its closest approach with Neptune. 
In the coplanar case, this closest approach occurs near the particle's conjunction with Neptune (i.e., when the Neptune lies on the line connecting the sun and the TNO). 
As a simple example, consider the 2:1 resonance, for which only one conjunction occurs per pattern period of the resonance.  
For an object in the 2:1 resonance on an inclined orbit with $\omega_{TNO} = 0\degr$ or $180\degr$, conjunctions near the particle's aphelion ($\phi=180^\circ$) or perihelion ($\phi=0^\circ$) occur in or near the plane of Neptune's orbit, and the closest approach to Neptune occurs near these conjunctions. 
The impact of kicks near these conjunctions is thus similar to the kicks experienced in the coplanar case, which drive libration about $180\degr$ \citep{peale1986,murray_dermott2000,murray-clay2005}.  
This geometry is shown in the left panel of Figure~\ref{f:conjunctions} for a 2:1 particle with $e=0.22$, $i=55^\circ$, $\omega_{TNO} = 0^\circ$ and an initial resonant angle $\phi=10^\circ$.
This particle has its closest approach to Neptune (and thus strongest kick) right after passing through perihelion at its ascending node and before conjunction with Neptune.
The net kick through closest approach is in the opposite direction of the particle's motion along its orbit, which has the effect of decreasing the particle's semimajor axis and increasing its mean motion. 
Thus the particle gets further along in its orbit before the next conjunction with Neptune, pulling conjunction even further away from its perihelion and moving the resonant angle further away from zero, just like in the coplanar case.

In contrast to the coplanar case, when $\omega_{TNO} = 90\degr$ or $270\degr$, conjunctions near the particle's perihelion or aphelion occur at maximum extents of its orbit above or below Neptune's orbital plane.  
The right panel of Figure~\ref{f:conjunctions} shows the same 2:1 particle as discussed above, but with $\omega_{TNO} = 90^\circ$. 
The particle again starts with $\phi=10^\circ$, which means its conjunction with Neptune occurs just after it passes through perihelion. 
However now, this conjunction is far above Neptune's orbital plane and very far from the particle's closest approach to Neptune, so the kick near conjunction is unimportant.
The closest approach happens slightly before the particle passes through its descending node, and the net kick from Neptune is downward and along the particle's direction of motion. 
This increases the particle's semimajor axis, which decreases its mean motion.
Thus Neptune catches up to the particle slightly by the time the particle reaches its next perihelion passage, moving conjunction back toward perihelion and decreasing the resonant angle back toward $\phi=0^\circ$, the opposite of expectations from the coplanar case.

For high particle inclinations, interactions near conjunction can have, relatively speaking, a smaller impact. 
This weakening of the direct interaction at conjunction can be significant enough that the indirect acceleration of the Sun by Neptune, which drives $\phi$ toward $0\degr$, can play a dominant role.
An initial exploration of the 2:1 resonance implies that there are critical eccentricity and inclination thresholds for this to occur and allow libration around $0\degr$, consistent with Figure~\ref{fig:res_center}, which show libration around $0\degr$ for only a portion of the parameter space explored by clones of 2020 VN$_{40}$.
For $n:1$ resonances in general, $(n-1)$ conjunctions occur per resonance pattern period, and libration of $\phi = 180\degr$ does not always correspond to conjunctions near the particle's apocenter.
However, when one of the conjunctions occurs at the particle's pericenter, $\phi = 0\degr$.  
Upon inspection, the averaged impact of all $(n-1)$ conjunctions exhibits a symmetry which allows the intuition from our 2:1 example to apply.
We comment that though only $n:1$ resonances have an indirect term in the disturbing potential, a cursory examination did uncover an example of libration around $\phi = 0\degr$ at high inclination for an $n:2$ resonance.  
We reserve a more detailed analysis of the circumstances under which eyehole libration arises, including an exploration of whether libration around $\phi=0\degr$ is dynamically distinct in $n:1$ resonances or merely enhanced by the indirect potential, for future work.

\section{Discussion}
\label{sec:discuss}

The \lido discovery 2020 VN$_{40}$ is likely sticking in the 10:1 resonance as part of its weakly scattering long-term evolution rather than being permanently trapped in the resonance.
The 201 test particles, including the nominal orbit and 200 clones, show predominantly resonant behavior on 30 Myr timescales.
The semi-major axis distribution of the clones in a Poincar\'e map of the resonance is also consistent with the object being currently resonant in the 10:1.
We use the 30~Myr timescale for a current classification of the object as a 10:1 resonator.
The dominant resonant island at the start of the simulation is the leading asymmetric island, including the nominal clone, though the clones move between islands repeatedly during the 30~Myr simulations.
Based on the 30~Myr simulations, the \lido project classifies this TNO (part of the characterized sample) as resonant in the 10:1, in the leading asymmetric island \citep{alexandersen2025}.

On Gyr timescales, over half of the test particles scatter away from the 10:1 resonance.
Half of the particles are resonant for a total of 0.3 Gyr during the 1 Gyr simulations.
Only 4 clones are resonant for more than 90\% of the Gyr simulation, with one resonant for 99\% of the simulation.
The most common resonant behavior over 1 Gyr is oscillation in the symmetric libration island, where 50\% of clones spend at least 0.12~Gyr.
The fraction of time spent in the two asymmetric islands is comparable to each other; 50\% of the clones are in the 90$\degr$ and 270$\degr$ islands for at least 0.080~Gyr and 0.072~Gyr respectively.
During the long-term evolution, approximately half of the clones also display periods of libration around $\phi=0\degr$.
Because of the large number of clones which scatter away from the resonance, and the low fraction of total time the particles spend in resonance even if they do not actively scatter, we consider this object a scattering sticking TNO.
The presence of many scattering sticking TNOs in the distant $n$:1 and $n$:2 resonances is expected based on previous work \citep[e.g.][]{Lykawka:2007}.
The discovery of this Myr but not Gyr stable 10:1 resonator is consistent with our expectations for large populations of TNOs in the distant resonances based on other $n$:1s and $n$:2s \citep[e.g.][]{alexandersen2016, Volk2018, Crompvoets2022}.
For some distant resonances, the number of detections implies that there might have been a significant population delivered in the past, with one possibility being a large scattering event in the past \citep{pike2015,Volk2018}.
However, a large population is expected simply from scattering and sticking from within the current distant TNO populations, and some of these sticks will last for $\sim$Gyr timescales like those seen for some of the clones of 2020 VN$_{40}$ \citep[e.g.][]{Yu2018}.
With a single detection in the 10:1, even from the characterized \lido survey, we cannot differentiate between these two population mechanisms for the 10:1 mean motion resonance.

In the 1~Gyr simulations, clones of 2020 VN$_{40}$ displayed a novel mode of libration within the the 10:1 resonance, libration around $\phi=0\degr$.
Our dynamical analysis of this behavior indicates that the large inclination  of the object ($i=33.407\degr$) is the driver of the new libration island.
The libration around $\phi=0\degr$ can be long-term stable and often creates a recurring pattern of libration around $\phi=180\degr$ and $\phi=0\degr$, with a smooth transition between these states (creating `eyehole' patterns in $\phi$ over time).
We determined that this behavior can occur generically for the $n$:1 resonances at large inclinations, and we are able to represent this behavior evolution using an averaged Hamiltonian.
In the clone simulations, the slow precession of $\omega_{\rm TNO}$ corresponds with the recurring pattern of libration around $\phi=180\degr$ and $\phi=0\degr$ that matches the patterns seen in the Hamiltonian curves.
The value of $\omega_{\rm TNO}$ affects the resonant evolution for large-$i$ objects because it affects where in a TNO's orbit it experiences its point of closest approach to Neptune.
As a result, the distribution of pericenter for $n$:1 resonators with mid-large inclinations is not necessarily restricted to exclude pericenter near $\phi=0\degr$ (longitudes near Neptune) like is expected for lower-inclination resonant populations. 

The \lido survey discovered 141 objects with $i>14\degr$, including the 10:1 resonator 2020 VN$_{40}$, which created a challenge for our typical models for distant $n$:1 resonators.
This object is temporarily resonant in the 10:1, and expected to be part of the resonant sticking population in the distant solar system.
2020 VN$_{40}$ revealed an unexpected resonant behavior in its long-term evolution, libration around $\phi=0\degr$.
Further investigation of additional \lido objects identified a few occurrences of similar behavior in a 7:1 resonator ($i=32.183\degr$) and a significant amount of this libration for a 5:1 resonator ($i=38.085\degr$).
Both of these objects are fainter than the characterization limit of \lido, but are indicative of the general prevalence of libration around $\phi=0\degr$ at high inclinations.
Additional investigation of the \lido objects is ongoing; however, the high fraction of unusual resonant behavior at mid-high inclinations indicates a need for meticulous tracking of distant objects, careful consideration of the on-sky distribution of $a>55$ resonant TNOs, and careful analysis of n-body integrations of these objects.
The toy model of three libration islands, two asymmetric and one symmetric as has previously been used successfully for $n$:1 resonant TNOs, was insufficient to describe and classify these objects, and our classification system required many iterations to correctly identify all of the periods of resonance.
For future work, we will also utilize the SBDyNT TNO classification tool\footnote{\url{https://github.com/small-body-dynamics/SBDynT}} \citep{Volk:2025}, which does not rely on knowing the possible libration centers of an object.
An unrestricted approach to identifying resonant TNOs will be critical as the TNO discoveries from Legacy Survey of Space and Time \citep[LSST,][]{Schwamb2023} are investigated and classified, as this survey will naturally discover many high-inclination objects.

\section{Acknowledgements}
The authors acknowledge the sacred nature of Maunakea and appreciate the opportunity to observe from the mountain. CFHT is operated by the National Research
Council (NRC) of Canada, the Institute National des Sciences de l’Universe of the Centre National de la Recherche Scientiﬁque (CNRS) of France, and the University of Hawaii, with \lido receiving additional access due to contributions from the Institute of Astronomy and Astrophysics, Academia Sinica, Taiwan. Data were produced and hosted at the Canadian Astronomy Data Centre; processing and analysis were performed using computing and storage capacity provided by the Canadian Advanced Network For Astronomy Research
(CANFAR), operated in partnership by the Canadian Astronomy Data Centre and The Digital Research Alliance of Canada with support from the National Research Council of Canada the Canadian Space Agency, CANARIE and the Canadian Foundation for Innovation.

Based on observations obtained at the international Gemini Observatory, a program of NSF NOIRLab, which is managed by the Association of Universities for Research in Astronomy (AURA) under a cooperative agreement with the U.S. National Science Foundation on behalf of the Gemini Observatory partnership: the U.S. National Science Foundation (United States), National Research Council (Canada), Agencia Nacional de Investigaci\'{o}n y Desarrollo (Chile), Ministerio de Ciencia, Tecnolog\'{i}a e Innovaci\'{o}n (Argentina), Minist\'{e}rio da Ci\^{e}ncia, Tecnologia, Inova\c{c}\~{o}es e Comunica\c{c}\~{o}es (Brazil), and Korea Astronomy and Space Science Institute (Republic of Korea), acquired through the Gemini Observatory Archive at NSF NOIRLab and processed using the Gemini IRAF package.
This paper includes data gathered with the 6.5 meter Magellan Telescopes located at Las Campanas Observatory, Chile.
This research is based in part on data collected at the Subaru Telescope, which is operated by the National Astronomical Observatory of Japan.

The authors wish to acknowledge the land on which they live and carry out their research.
 Center for Astrophysics | Harvard \& Smithsonian is located on the traditional and ancestral land of the Massachusett, the original inhabitants of what is now known as Boston and Cambridge. We pay respect to the people of the Massachusett Tribe, past and present, and honor the land itself which remains sacred to the Massachusett People.
University of Regina is on Canadian Treaty 4 land, which is the territories of the n\^{e}hiyawak, Anih\v{s}in\={a}p\={e}k, Dakota, Lakota, and Nakoda, and the homeland of the M\'{e}tis/Michif Nation.

REP, MA, and CC acknowledge NASA Solar System Observations grant 80NSSC21K0289.
REP, KV, RMC, and AHR acknowledge NASA Emerging Worlds grant 80NSSC21K0376 and NASA Solar System Observations grant 80NSSC23K0680.
KV acknowledges additional support from NASA grants 80NSSC23K0886 80NSSC23K1169.
CC was supported in part by a Massachusetts Space Grant Consortium (MASGC) Award.
SML was supported in part by NSERC Discovery Grant RGPIN-2020-04111.

\facilities{CFHT, CANFAR, Magellan} 
\software{This research was made possible by the open-source projects 
\texttt{find\_orb} \citep{Gray2022},
\texttt{REBOUND} \citep{rein2012},
\texttt{WHFast} \citep{rein2015},
the updated version of the CFEPS Moving Object detection Pipeline \citep{Petit2004} used by OSSOS,
\texttt{MegaPipe} \citep{Gwyn2008}}.

\end{document}